\documentclass[preprint,aps,epsf]{revtex4}
\usepackage{graphicx}
\def\s{\sigma}

\def\up{\uparrow}
\def\dd{\downarrow}
\def\las{\langle}
\def\ras{\rangle}

\begin{document}
\title{Temperature and voltage dependence of magnetic barrier junctions
with a nonmagnetic spacer}
\author{Ali A. Shokri}
\altaffiliation{E-mail: aashokri@mehr.sharif.edu}
\affiliation{Department of Physics, Sharif University of
Technology, 11365-9161, Tehran, Iran}
\author{Alireza Saf\mbox{}farzadeh}
\altaffiliation{Corresponding author. E-mail:
a-saffar@tehran.pnu.ac.ir} \affiliation{Department of Physics,
Tehran Payame Noor University,\\ Fallahpour St.,
Nej\mbox{}atollahi St., Tehran, Iran}
\date{\today}

\begin{abstract}
The temperature and voltage dependence of spin transport is
theoretically investigated in a new type of magnetic tunnel
junction, which consists of two ferromagnetic outer electrodes
separated by a ferromagnetic barrier and a nonmagnetic (NM)
metallic spacer. The effect of spin fluctuation in magnetic
barrier, which plays an important role at finite temperature, is
included by taking the mean-field approximation. It is found that,
the tunnel magnetoresistance (TMR) and the electron-spin
polarization depend strongly on the temperature and the applied
voltage. The TMR and spin polarization at different temperatures
show an oscillatory behavior as a function of the NM spacer
thickness. Also, the amplitude of these oscillations is regularly
reduced when the temperature increases. The maximum TMR value,
varies approximately from 270$\%$ in reverse bias (at $T$=0 K) to
25$\%$ in forward bias (at $T\geq T_C$).
\end{abstract}
\maketitle
\newpage

\section{Introduction}
Spin-dependent transport in the magnetic multilayers and magnetic
tunnel junctions (MTJs) has created the new field of spin
electronics, where the electron spin and charge are used to design
new devices, such as spin transistors \cite{Mons} and memory cells
in magnetic random access memories \cite{Mo,Parkin}. The
performance of such magnetoelectronic devices depends critically
on the structure of magnetic junctions and the degree of spin
filtering in the ferromagnetic (FM) components. Among the
structures, a MTJ which consists of two FM metallic layers
(electrodes) separated by a thin insulator (I) layer, has
attracted much attention. On the other hand, the spin-polarized
resonant-tunneling effects in some of the systems, have shown
that, the insertion of a NM metallic layer between the tunnel
barrier and the FM electrode is especially important for the
development of highly functional spin electronic systems
\cite{Sun,Lec11,Lec22,Lec33,Zhang}.

The recent experimental results on the FM/I/NM/FM structure have
shown that, due to the decoherence tunneling of electrons, the TMR
reduces, when the thickness of the NM spacer increases
\cite{Mood1}. However, in another experiment, using a single
crystal for the NM/FM electrode, an oscillatory behavior for the
TMR has been observed \cite{Yuasa}. These oscillations have been
attributed to the quantum well states formed in the NM spacer
layer. Since in each MTJ the TMR strongly depends on the current
spin polarization, a possible way to fabricate MTJs with more
highly spin polarization, is to apply a ferromagnetic
semiconductor (FMS) barrier instead of nonmagnetic insulator
layer.

During recent years, the spin transport using FMS layers, has been
studied both experimentally \cite{Lec1,Lec2} and theoretically
\cite{Worl,Wilc,Saffar1,Saffar2,Shok} in MTJs. When a FMS layer
(magnetic barrier) is used in tunnel structures, due to the spin
splitting of the FMS conduction band at below the Curie temperature
($T<T_C$), the tunneling electrons see the spin-dependent barrier
heights. Therefore, the probability of tunneling for one spin channel
will be much larger than the other, and a highly spin-polarized
current may result \cite{Mood2,Mood3}.

In the previous paper, we studied the effects of a FMS layer
in FM/FMS/NM/FM tunnel junction, at $T$=0 K \cite{Shok}. We
showed that, for particular thicknesses of the NM layer, due to
the  magnetization of the FMS layer and also, the formation of the
resonance states in the NM layer, very large TMR effects can be
obtained.

In this paper, using the transfer matrix method and the nearly
free electron approximation for spin-dependent transport in the
FM/FMS/NM/FM tunnel junction, the effects of the temperature, the
thickness of the NM layer and the applied bias on the TMR and spin
polarization of the tunneling current are investigated. We assume
that the electron wave vector parallel to the interfaces and the
spin direction of the electron are conserved in the tunneling
process through the whole system. Therefore, there is no spin-flip
process in this study.

The structure of this paper is as follows. In section 2, the model
and method is briefly described. In section 3, the numerical results
for the TMR and spin polarization of the tunnel current in a typical
tunnel junction are presented and discussed. The results are summarized
in section 4.

\section{Method of calculation}
Consider a MTJ composed of two semi-infinite FM electrodes
separated by a FMS layer and a NM metallic spacer, in the presence
of an applied bias $V_a$, as shown in Fig. 1. In the FM/FMS/NM/FM
structure, the FMS layer acts as a spin filter and the NM layer as
a quantum well. For simplicity, we assume the two FM electrodes
are made of the same material. In the absence of any kind of
scattering center for electrons, the motion along the $x$-axis is
decoupled from that of the $y$-$z$ plane. Therefore, in the
framework of the effective mass approximation, the longitudinal
part of the one-electron Hamiltonian can be written as
\begin{equation}\label{H}
H_x=-\frac{\hbar^2}{2m_j^*}\frac{d^2}{dx^2}+U_j(x)
-{\bf h}_j\cdot{\bf\sigma}+V^\sigma,
\end{equation}
where $m^*_j$ ($j$=1-4) is the electron effective mass in the $j$th
layer, and
\begin{equation}\label{U}
U_j(x)=\left\{\begin{array}{cc}
0, & x<0 \ ,\\
E_{\rm FL}+\phi-eV_ax/t_{\rm bar}, & 0<x<t_{\rm bar}\  ,\\
-eV_a, & t_{\rm bar}<x<t_{\rm bar}+t_{\rm NM}\  ,\\
-eV_a, & x>t_{\rm bar}+t_{\rm NM}\  ,\\
\end{array}\right.
\end{equation}
where $E_{\rm FL}$ is the Fermi energy in the left electrode.
$-{\bf h}_j\cdot{\bf\s}$ is the internal exchange energy where
${\bf h}_j$ is the molecular field in the $j$th FM electrode and
${\bf\s}$ is the conventional Pauli spin operator. The last term
in Eq. (\ref{H}) is a spin-dependent potential and denotes the
$s-f$ exchange coupling between the spin of tunneling electrons
and the localized $f$ spins in the FMS layer. Within the
mean-field approximation, $V^\sigma$ is proportional to the
thermal average of the $f$ spins, $\langle S_z\rangle$ (a 7/2
Brillouin function), and can be written as $V^{\sigma}=-I\s\langle
S_z\rangle$. Here, $I$ is the $s-f$ exchange constant in the
magnetic barrier, and $\sigma=\pm1$ (``+" for spin-up electron and
``-'' for spin down one).

In order to investigate the spin transport properties in the
present structure, we calculate the spin-dependent transmission
coefficients, $T_\sigma(E,V_a,T)$ using the transfer matrix method
\cite {Shok}. We should note that, the transmission coefficients
depend on the energy $E (=E_x+E_\|)$, the applied bias $V_a$, the
temperature $T$, the alignment of magnetizations in magnetic
layers as well as spin orientation. Therefore, the temperature and
voltage dependence of current density for spin $\s$ electron, in
the parallel (antiparallel) alignment, can be determined as
\cite{Duke}:
\begin{equation}\label{J}
J^{\rm p(ap)}_\sigma(T,V_a)=\frac{em^*_1}{4\pi^2\hbar^3}
\int_{0}^{\infty}dE_x\int_{0}^{\infty}dE_\|\left[f(E)-f(E+eV_a)
\right]T^{\rm p(ap)}_\sigma(E,V_a,T),
\end{equation}
where $f(E)=[1+\exp(E-E_F)/k_BT]^{-1}$ is the equilibrium
Fermi-Dirac distribution at temperature $T$. The degree of spin
polarization is defined as the difference between the spin-up and
spin-down current densities:
\begin{equation}\label{p}
P=\frac{J^{\rm p}_\up-J^{\rm p}_\dd}{J^{\rm p}_\up+J^{\rm p}_\dd}.
\end{equation}
On the other hand, using the current densities in the parallel and
antiparallel configurations, the TMR can be described quantitatively
by the relative current change as
\begin{equation}\label{tmr}
\mbox{TMR}=\frac{(J_{\up}^{\rm p}+J_{\dd}^{\rm p})-(J_{\up}^{\rm
ap} +J_{\dd}^{\rm ap})}{J_{\up}^{\rm ap}+J_{\dd}^{\rm ap}}.
\end{equation}

In our considered system, the magnetization direction of the left
FM electrode and the FMS layer stays fixed, but the right FM
electrode is free and may be switched back and forth by an
external magnetic field, as shown in Fig. 1. Thus, the spin
transport is modulated by the magnetic alignment of the right FM
electrode.

\section{Numerical results and discussion}
Numerical calculations have been carried out to investigate the
effects of temperature and applied voltage on spin transport in
Fe/EuS/Au/Fe structure as a typical MTJ. We have chosen Fe and EuS
because they have cubic structures and the lattice mismatch is
very small \cite{Dem}. The relevant parameters for the Fermi
energy and the internal exchange energy in the FM electrodes are
chosen as $E_{\rm F}$=2.62 eV and $h_0$=1.96 eV, which correspond
to itinerant $d$ electrons in Fe \cite{Stearns}. The suitable
parameters for EuS as a magnetic barrier are $T_{\rm C}$=16.5 K
\cite{Baum}, $S$=7/2, $I$=0.1 eV \cite{Nolt}, $\phi$=1.94 eV
\cite{Saffar2} as a symmetric barrier height, and $t_{\rm
bar}$=1.3788 nm which corresponds to four monolayer (ML)
thicknesses of EuS$\las 111\ras$ \cite{Maug}. This thickness is
constant in all calculations. The Fermi energy in the Au layer is
$E_{\rm F,NM}=5.51$ eV \cite{Kittel}, and the thickness of this
layer, $t_{\rm NM}$, will be determined in terms of the interlayer
distance of Au$\las 111\ras$ which is equal to 1ML=0.2355 nm. The
effective mass of all electrons for the structure are taken as the
free electron mass $m_e$.

In Figs. 2 and 3, we have shown the tunneling current densities
and the TMR as a function of the applied voltage at different
temperatures, when $t_{\rm NM}=3\rm ML$. At low voltages the
current densities vary linearly, whereas, with increasing the
applied voltage, the effective width of the magnetic barrier
becomes narrower and a nearly parabolic dependence of current on
the voltage appears. These curves are typical of what is expected
from Brinkman model \cite{Brink} describing the tunneling free
electrons. At high temperatures $T>T_C$, there is no spin
splitting of the FMS conduction band, so the magnetic
barrier acts as a nonmagnetic insulator. In this case, the
difference between current densities in both parallel and
antiparallel configurations, and hence, the TMR effect is only due
to the FM electrodes. Thus, in such temperatures the TMR has low
values; see Fig. 3 at $T$=1.2 $T_C$. As the temperature decreases
from $T_C$, the barrier height for spin-up electrons is lowered,
while it is raised for spin-down ones; thus, for a fixed
applied voltage, with decreasing the temperature, the difference
between current densities in the parallel and antiparallel
configurations and then the TMR, increases. Both figures confirm
that when the current densities in both configurations cross each
other, the sign of TMR be reversed. Fig. 3 also shows that the
highest value of TMR (about 270$\%$) is obtained in reverse bias
and at zero temperature. This value reduces to nearly 25$\%$ in
forward bias and at $T>T_C$.

In the previous work, we showed that, with continuous variation of
the NM layer thickness, the TMR oscillates with a short period
equal to $\pi/k_F=0.26$ nm, where $k_F$ is the Fermi wave vector
in the Au layer. In the present study, we have investigated the
dependence of TMR on the NM spacer thickness at different temperatures,
when $t_{\rm NM}$ changes in monolayer steps ($d$), as it is shown
in Fig. 4. It is interesting to note that, at all temperatures the
TMR first increases in the first three monolayers and after that,
it becomes less and starts to oscillate with a long period of
approximately 2.5 nm. The value of this modification of
oscillation period can be obtained from $\pi/|k_F-n\pi/d|$ (with
$n$ an integer), which is called \emph{aliasing} or \emph{Vernier}
effect. This effect is well understood from the theory of Fourier
analysis by taking into account the discrete variation of the
NM spacer thickness \cite{Bruno91,Coe91}.

With decreasing the temperature from $T\geq T_{\rm C}$, the
amplitude of oscillations increases, whereas, with increasing the
NM layer thickness, the amplitude decreases. The origin of TMR
oscillations is related to the quantum well states, formed in the
NM spacer \cite{Shok}.

In Fig. 5 we have displayed the spin polarization of tunneling
electrons as a function of normalized temperature $T/T_C$, for
several monolayer thicknesses  $t_{\rm NM}$. At the temperatures
that the FMS layer acts as a nonmagnetic insulator, the spin
filtering effect in the tunnel currents is only due to the FM
electrodes. Therefore, the electron spin polarization is not very
high. However, at very low temperatures, the FMS layer strongly
affects the TMR and spin polarization. The highest value of the
spin polarization can reach 97\% at zero temperature, when 3ML are
used for the NM layer. With increasing the temperature, this value
reduces to 60\%, at $T  \geq T_C$. The thickness dependence of
spin polarization has also shown in the inset of Fig. 5. It is
obvious that, by variation of the spacer thickness, the spin
polarization oscillates with a period which is equal to the
oscillation period of the TMR.

It is necessary to point out that, the magnetization of FM
electrodes generally depends on the temperature \cite{Mood4}. For
this reason, we studied the effect of temperature on
the magnetization of FM electrodes, by considering a term
proportional to $T^{3/2}$ for this quantity. This term which has
been experimentally confirmed, is applicable for surface
magnetization. Since tunneling phenomenon is a surface-sensitive
process, one can consider such temperature dependence for
magnetization of FM electrodes. Owing to the proportionality
between exchange field $|{\bf h}_j|$ and the surface magnetization
of the FM electrodes, we can write $|{\bf h}_j(T)|\propto T^{3/2}$.
The obtained results showed that, the effect of temperature variation
on the magnetization of electrodes, in comparison with this effect in
the FMS layer at $T<T_{\rm C}$, is very negligible, so we assumed that
the FM electrodes are in completely ferromagnetic case.

Therefore, the calculated results show that, the temperature dependent
spin transport is a result of the ferromagnetic phase of the FMS tunnel
barrier and by adjusting the temperature, applied voltage and the
thickness of NM layer, one can reach high values for the tunneling
spin-polarization and TMR.

\section{Concluding remarks}
The temperature and voltage dependence of TMR and the spin
polarization were theoretically investigated, in a new type of MTJ
based on the nearly free-electron model and the transfer matrix
method. Numerical results indicate that in the Fe/EuS/Au/Fe
structure, due to the quantum well states and the magnetic barrier,
there exist more than 260\% TMR effect and about 97\% spin
polarization for tunneling electrons. At fixed temperature and
voltage, the TMR has an oscillatory behavior as a function of the
NM layer thickness. Because of the strong filtering effect of the
magnetic barrier, the oscillations persist up to very large
thicknesses and the period of these oscillations at all temperatures,
along a fixed $\las hkl\ras$ direction inside the Au layer, is equal.

In this study, we used of a low temperature FMS as a magnetic barrier.
However, due to the recent predictions of the room-temperature FMSs
\cite{Dietl}, the present results may have potential utility for
designing new spin electronic devices such as resonant-tunneling
spin transistor and digital storage technology \cite{Yuasa}.

\newpage
\begin{figure}
\centering \resizebox{0.6\textwidth}{0.4\textheight}
{\includegraphics{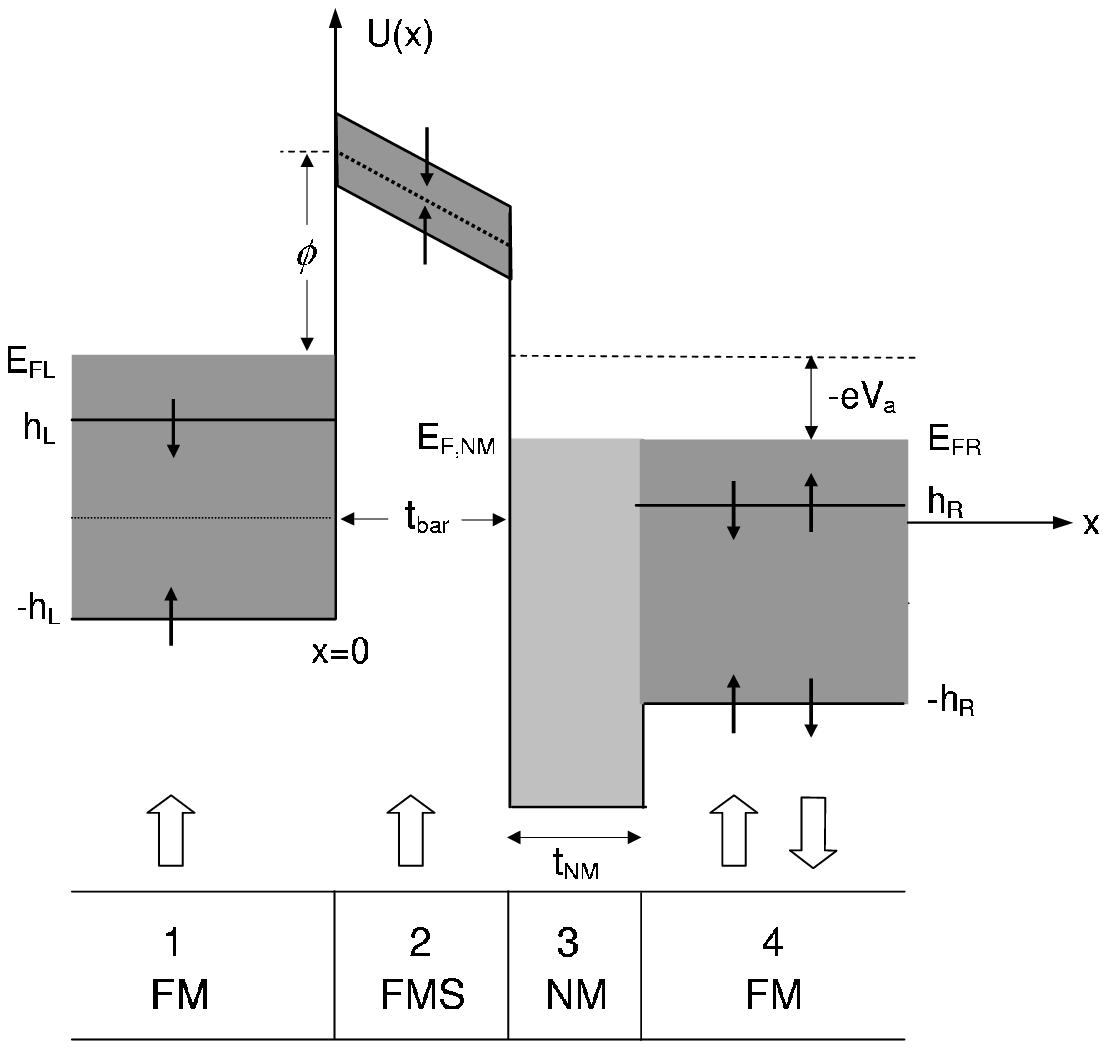}} \caption{Spin-dependent potential
profile for FM/FMS/NM/FM magnetic tunnel junction under forward
bias $V_a$. In the FMS layer, the dashed line represents the
bottom of the conduction band at $T\geq T_C$ and the thin arrows
indicate the bottom of the conduction band for spin-up and
spin-down electrons at $T<T_C$. The zero of energy is taken at
the middle of bottoms for majority-spin band and minority-spin
one in the left FM electrode.}
\end{figure}
\begin{figure}
\centering\resizebox{0.6\textwidth}{0.4\textheight}
{\includegraphics{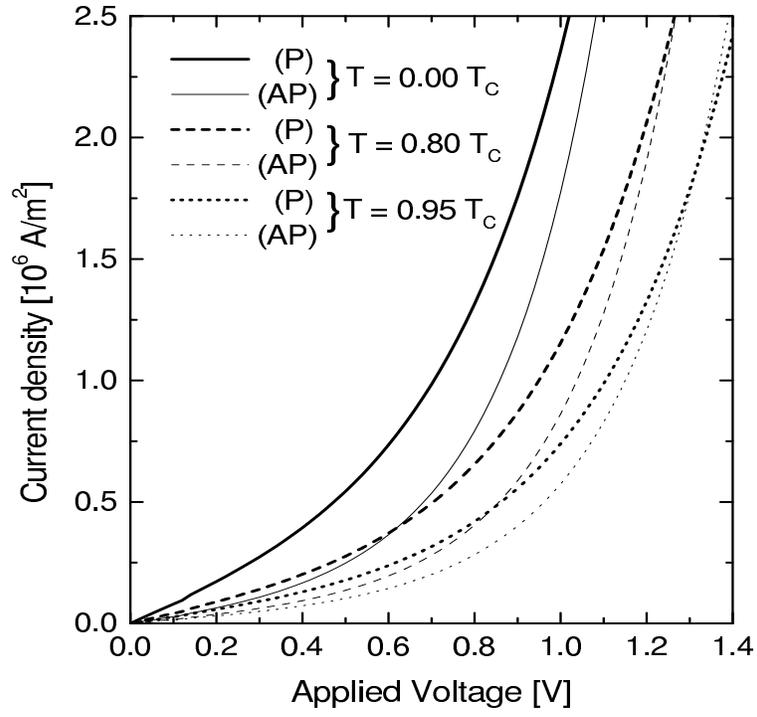}} \caption{Variations of the tunneling
current densities as a function of applied voltage for different
temperatures in the parallel (P) and antiparallel (AP)
configurations.}
\end{figure}
\begin{figure}
\centering \resizebox{0.5\textwidth}{0.5\textheight}
{\includegraphics{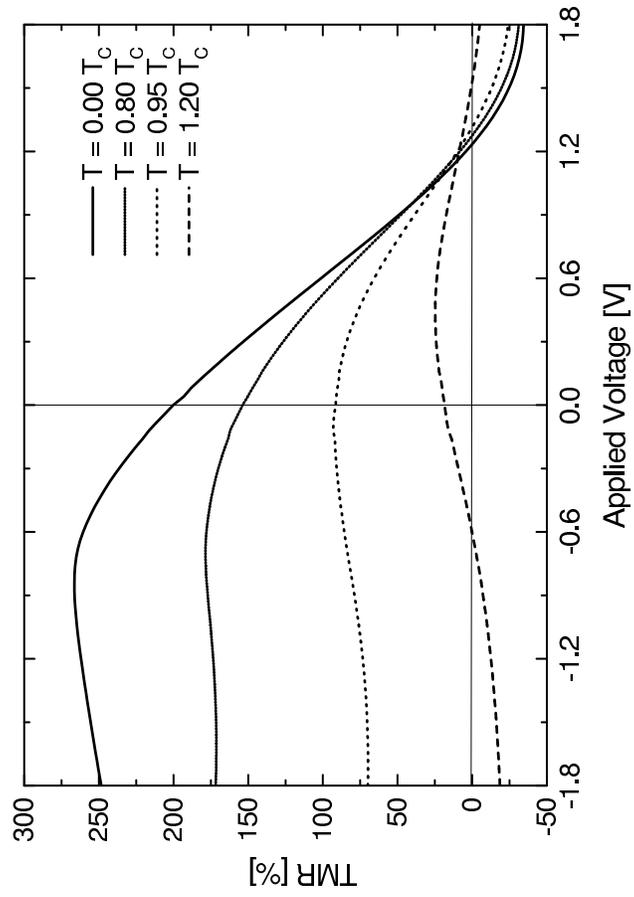}} \caption{Dependence of the TMR on
the applied voltage at different temperatures for $t_{\rm
NM}=$3ML.}
\end{figure}
\begin{figure}
\centering \resizebox{0.5\textwidth}{0.5\textheight}
{\includegraphics{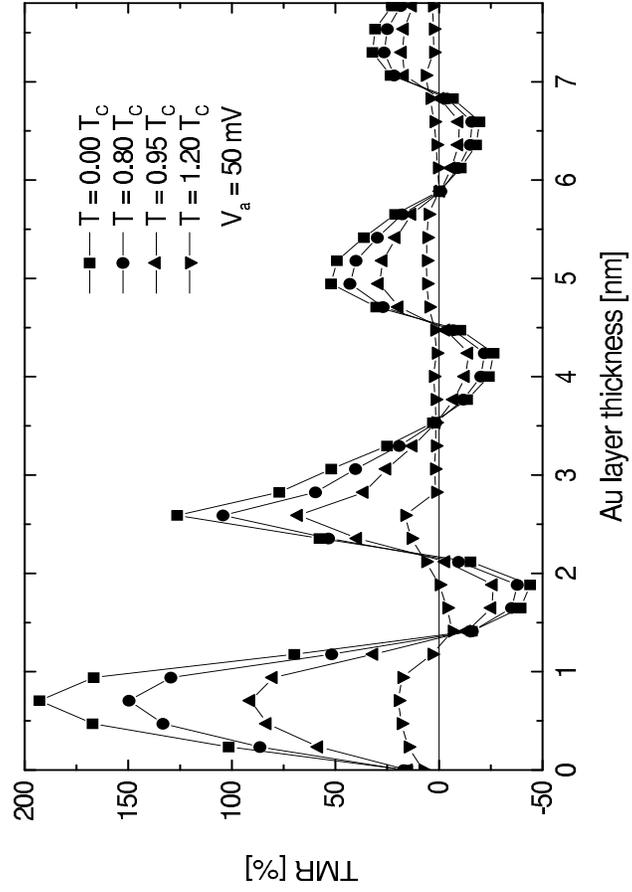}} \caption{Dependence of the TMR as a
function of the Au layer thickness at different temperatures.}
\end{figure}
\begin{figure}
\centering \resizebox{0.5\textwidth}{0.5\textheight}
{\includegraphics{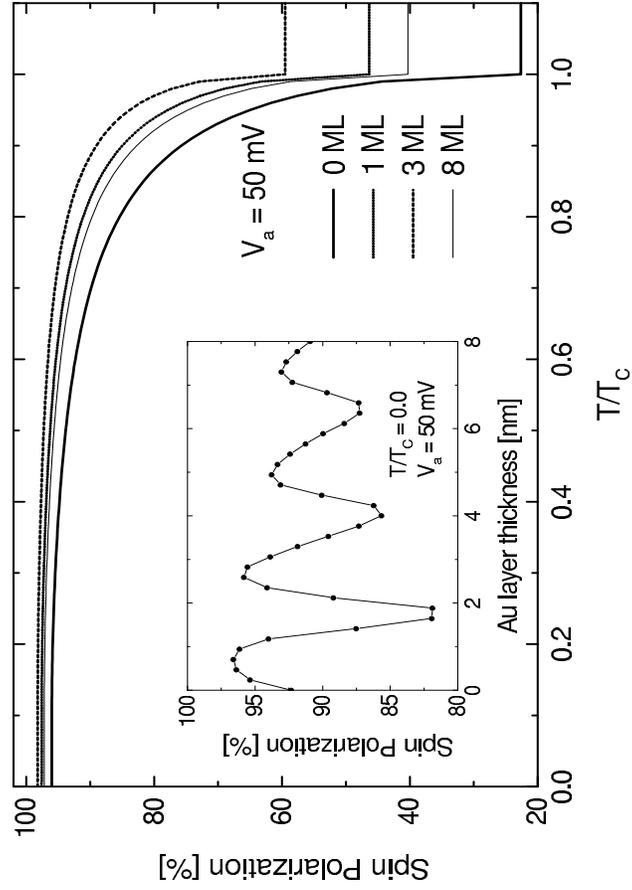}} \caption{Dependence of the spin
polarization as a function of normalized temperature for several
NM multilayers. Inset demonstrates the dependence of spin
polarization on the Au layer thickness.}
\end{figure}

\end{document}